\documentstyle[aps,preprint]{revtex}
\draft \tighten
\input epsf

\begin{document}

\def\pbg{\hbox{${\mathbf \pi\beta\gamma}$}}
\def\tsigma{\tilde{\sigma}}
\def\lesssim{\mathrel{\hbox{\rlap{\hbox{\lower4pt\hbox{$\sim$}}}\hbox{$<$}}}}
\def\gtrsim{\mathrel{\hbox{\rlap{\hbox{\lower4pt\hbox{$\sim$}}}\hbox{$>$}}}}
% \twocolumn[\hsize\textwidth\columnwidth\hsize\csname
% @twocolumnfalse\endcsname
\preprint{CITA-2001-19}
\date{July 26, 2001}

\title{Tachyonic Preheating\footnote{Invited talk at the Conference PASCOS2001}}

\author{Lev Kofman}

\address{Canadian Institute for Theoretical Astrophysics,
University of Toronto, ON M5S 3H8
\\E-mail: kofman@cita.utoronto.ca
}

\maketitle
\begin{abstract}
I review the theory of preheating after inflation, focussing
on the recently found tachyonic preheating in the theories with 
spontaneous symmetry breaking. This 
occurs due to the tachyonic instability of the scalar field near the top of its
effective potential. Contrary to the common expectation,
tachyonic instability  converts most of the energy into that of colliding
classical waves very rapidly, within a single oscillation.
Efficient tachyonic preheating is typical
for  the hybrid inflationary scenario, including SUSY motivated and brane inflation  models.
\end{abstract}

\section{Preheating after Inflation}

The best-fit universe is  the  uniform and homogeneous flat
 $\Omega_{tot}=1$  Friedmann-Lemetre
expanding    universe with gaussian scale-free  metric  fluctuations.
The model is in remarkable agreement with  the observations  of the
CMB anisotropy, the large scale structure
 of the universe and the tests of the global geometry.

Inflationary theory  suggests that the expansion of the
universe was preceded not by  the Big Bang singularity, but rather
by  the exponentially fast expansion of the universe when
four dimensional geometry of the universe is close to the nonsingular
De Sitter geometry.
 Almost De Sitter geometry (where the Hubble parameter is slow varying with time,
$\dot H \ll H^2$)
can be realized by gravity of the matter with
the vacuum-like equation of state $p\approx -\rho$, where  $p$ and $\rho$ are
 the pressure and energy density.
There are many specific realizations of this equation of state,
motivated by the different aspects of the elementary particle  physics.
The vacuum-like  equation of state
can be provided by the out-of-equilibrium   scalar fields  or by the other effects
(like the quantum gravity mechanisms) which behave like an effective  scalar field.

Although scalar fields are not yet discovered experimentally, they are the vital ingredients
of the high energy physics  theories, and a plethora of scalar fields exists
in the supergarvity and superstring theories.
Fundamental M-theory  should encompass both supergravity
and string theory.
At present the low-energy phenomenology is described by the $N=1$ $d=4$
supergravity.
Its  Lagrangian
begins with the scalar fields terms
\begin{equation}\label{lag}
e^{-1}{\cal L}=-{1 \over 2}M_P^2 R-
\partial_{\mu}  \Phi^i\partial^{\mu}  \Phi_i+
 e^K\left({\cal D}^iW {\cal D}_iW  -3 {{WW^*} \over M_P^2} \right)+\,...
\end{equation}
where
 a scalar field
$\Phi^i$ is the complex conjugate of $\Phi_i$.
 Some preferable
choices of the K\"{a}hler  potentials $K$, superpotentials $W$ and Yang--Mills
couplings hopefully will be selected at the level of the  fundamental theory.
Until the fundamental theory of all interactions is well understood,
one may try to address the issues of the early universe  cosmology
in the context of the most general phenomenological $N=1$
supergravity--Yang--Mills--matter theory. This, in fact,
was the case during the last decade.

At first glance it seems hard to gain  useful insights or predictions
from a 
loose pool of inflationary models embedded in (\ref{lag}).
 Fortunately, theory of inflation, independently on its
concrete model,   provides  us with several  universal    mechanisms
which make it immensely successful  as the theory of the
 initial conditions for the observable Friedmann-Lemetre universe.
These mechanisms are related to the properties of the DeSitter solution in 
General Relativity
and to the properties of classical and quantum fields in the  background
of De Sitter space-time.
Remarkably, these properties are not directly related to the microscopic
physics behind the 
inflation. This is not the  first time when we can successfully do
cosmology at the phenomenological level without the knowledge of the microscopic
physics.
For example, celebrated Cold Dark Matter theory of the Large Scale Structure 
of the universe
relies  on the simple assumption that there is  a nonbaryonic invisible dark matter
component with the "dust-like" equation of state $p_{cdm} \approx 0$. This is enough to
develop the theory of the structure formation even
 without understanding  the microscopic origin of dark matter.

There are  several  major predictions  of  inflation:\\
$\bullet$  Inflation erases   pre-inflationary  classical inhomogeneities and  entropy.\\
$\bullet$ The total mass density in the universe $\Omega_{tot}=1$.\\
$\bullet$ Vacuum quantum fluctuations of the scalar field(s) generate
    almost scale free gaussian scalar metric perturbations.\\
$\bullet$ Vacuum quantum fluctuations of gravitational waves
generate  almost scale free gaussian tensor metric perturbations.\\
$\bullet$ All the particles in the universe are created from the decay of inflaton energy $\rho$
in the process of (p)reheating  after inflation. \\
The ratio of amplitudes of scalar and tensor modes, the spectral index  of scalar and
 tensor modes,
and the character of  (p)reheating after inflation can be  model dependent.

In this contribution I will review the theory
 of preheating after inflation, concentrating on the recently found
tachyonic preheating \cite{GBFKLT,FKL}.

According to  the inflationary scenario,
 the Universe initially expands quasi-exponentially
in a vacuum-like (DeSitter) state without entropy or particles.
At the stage of inflation, all energy is contained
 in a classical slowly moving  fields $\Phi$ in the inflaton sector.
The last term of the 1st
line of (\ref{lag}) is the scalar potential $V(\Phi_i)$.
The equations of motion
 based on the first line
should describe inflation, which is a challenging problem by itself.
The Lagrangian  (\ref{lag})
contains also  other fields
which give subdominant contributions to gravity.
The Friedmann equation for the scale factor $a(t)$ and
the  equation for $\Phi(t)$
determine the evolution of the background fields.

In  the chaotic inflation models, soon after the end of inflation,
an almost homogeneous inflaton field $\Phi(t)$   coherently
oscillates with a very large  amplitude of the order of the Planck mass $M_P$
around the minimum of its potential $V(\Phi)$. This scalar field can be considered as
a coherent superposition of inflatons  with zero momenta.
 The amplitude of  oscillations  gradually
decreases not only because of the expansion of the
universe, but also because energy is transferred to particles
created by the oscillating field.
At this stage
we shall recall the rest of the fundamental Lagrangian
  which
includes all the fields interacting with inflaton.
These interactions
  lead  to the  creation of many ultra-relativistic
particles from the inflaton.
Gradually, the inflaton field decays and transfers
all of its energy    to the created
particles.
In this scenario
 all the matter constituting the universe
is created from this process of reheating.
If the creation of particles is sufficiently slow,
   the  particles would   simultaneously
 interact with each other and come to a state of thermal equilibrium
at  the reheating temperature $T_R$.
Typically particle production from coherently oscillating
inflatons occurs not  in the non-perturbative
regime of preheating \cite{KLS}.

Indeed, let us consider a simple toy  model
of chaotic inflation with the  quadratic
potential $V(\Phi)={1 \over 2} m_{\phi} \Phi^2$ and
${\cal L}_{int} =-{1 \over 2} g^2 \Phi^2 \chi^2$
 describing the interaction between the
 inflatons $\Phi=\Phi_1$ and other massless
Bose particles $\chi=\Phi_2$.
We can consider quantum fluctuations of the field  $\chi$ 
interacting with the classical homogeneous
 background field $\Phi(t)$.
The quantum scalar field $\hat \chi$ in a flat FRW background has
 the eigenfunctions $\chi_{k}(t)\, e^{ -i{{\bf k}}{{\bf x}}}$
 with comoving momentum ${\bf k}$.
 The temporal part of the eigenfunction   obeys the equation
\begin{equation}\label{res}
\ddot \chi_k + 3{{\dot a}\over a}\dot \chi_k + {\left(
{{\bf k}^2\over a^2}   - \xi R + g^2\phi^2 \right)} \chi_k = 0  \label{2}
\end{equation}
with vacuum-like initial conditions: $ \chi_k \simeq {e^{ -ikt} \over
 \sqrt{2k}}$
in the far past. The
coupling to the curvature $\xi R$ will not be important
in the presence of the interaction (but would lead to gravitational
preheating in the absence of the interaction).
In this model, the inflaton field $\Phi(t)$   coherently
oscillates as $\Phi(t) \approx \bar \Phi(t) ~
 \sin{\left( m_{\phi}t \right)}$, with the amplitude
 $ \bar \Phi(t)= {M_p \over \sqrt{3\pi}}\cdot{1  \over m_{\phi}t}$
 decreasing as the universe expands.

Equation (\ref{res}) describes excitation of the quantum fluctuation $\chi_k$.
At first glance the effect of particle creation  $n_{\chi}$ 
could be treated perturbatively with respect to the small coupling $g^2$.
However, 
the smallness of $g^2$ alone does not necessarily lead to the
perturbative approach to describe
the excitation of $\chi_k$ modes.
To check whether the interaction term $g^2\phi^2$ in eq. (\ref{2})
is perturbative or not, we have  to use
a new time variable $z=mt$ and the essential
dimensionless coupling parameter $q= {{g^2 \Phi^2} \over{ m^2}}$.
 Scalar metric fluctuations in this model
are compatible with cosmology if  the  inflaton mass is
 $m_{\phi}  \simeq 10^{-6}  M_p$; therefore,
 typically $q \simeq 10^{10} g^2 \gg 1$
for not negligibly small $g^2$.
In fact, a consistent setting for the problem of $\chi$-particle
creation from the $\phi$-inflaton requires $q \gg 1$
 even without additional assumptions about $g^2$. It is known that
if we have two scalars $\phi$ and $\chi$, then
the latest stage of inflation will be driven by the lightest
scalar. The square of the effective mass of the $\chi$-field includes a term
$g^2\phi^2$. Inflation is driven by the $\phi$-field if its square mass $m^2$
is smaller than $g^2\phi^2$. This leads to the condition $q \gg 1$.

Solutions of the equation (\ref{res}) for $q \gg 1$ are unstable for
 a range of $k$
and $\chi$ particles are created  exponentially fast.
 Indeed, suppose that there is no expansion of the universe.
In this case background oscillations are harmonic, 
$\Phi(t) \sim  \sin{\left( m_{\phi}t \right)}$, and  equation (\ref{res})
is reduced to the Mathieu equation. Solutions of this equation are
exponentially unstable within the set of resonance bands, $\chi_k \sim e^{\mu_k m_{\phi}t}$,
where the characteristic exponent$\mu_k$ depends on $(k, q)$.
In realistic case of the expanding universe, parameter $q$ is time dependent.
For the broad resonance case $q \gg 1$ this parameter jumps over a number of
instability bands within a single background oscillation, so the concept of
stability/instability bands is inapplicable here. Parametric
resonance in this case, described by  (\ref{res}), is a stochastic process \cite{KLS97}.
 In the regime of stochastic resonance particle are created exponentially fast as 
 $n_k \sim e^{2\int dt \mu_k}$. Characteristic exponent here is also
 stochastic   with
typical value $\bar \mu_k \sim 0.1$ in the resonance  momentum range $0 < k< q^{1/4}m$.

Due to the copious particle creation from the background field, 
very soon, within few dozens of the background oscillations, we have to
take into accound backreaction of the created particles.
Unfortunately,  known theoretical approaches to include 
backreaction effects, 
 like the Hartree approximation, are not enough in this situation.
First  fully nonlinear lattice simulation of the preheating \cite{lattice} revealed that 
the leading backreaction effect is rescattering of $\chi$ particles 
on the background inlatons,
 $\chi_{\vec k} \Phi_0 \to \chi_{\vec k'} \Phi_{{\vec k}-{\vec k'}}$.
Lattice simulations demonstrated that there is a first, resonant stage
of preheating, where  initial  excitation
of $\chi$ field occurs exponentially fast as it is expected from the linear  theory of 
equation (\ref{res}). Then the system enters the nonlinear stage 
of preheating where classical waves of $\chi$ and $\Phi$ fields 
are rescattering and gradually relaxing towards equilibrium.

Different in technical details, but qualitatively similar theory was developed 
for other types of the   chaotic inflation potentials, e.g. see \cite{GKLS}
for $V(\Phi)= { 1 \over 4} \lambda \Phi^4$ theory.

\section{Preheating in Hybrid Inflation}

Another popular class of inflationary models --
hybrid inflation -- involve multiple scalar fields $\Phi_i$ in the inflaton
sector \cite{Hybrid}. In particular, hybrid inflation 
can be realized in supergravity for certain choice of superpoteantial $W$
\cite{Stewart}, and in the theory of brane inflation \cite{brane}.

Previous studies of preheating in hybrid models
were concentrated on particle creation by parametric resonance that may
occur when homogeneous background fields oscillate around the minimum of
the potential \cite{hybrid1,hybrid2}.
Technically the linear fluctuations were considered around 
the time-dependent homogeneous background fields.
 Such parametric resonance may or may not be strong
depending on the coupling parameters. However,
 we  recently found \cite{GBFKLT} that there is strong preheating in
hybrid inflation, but its character is quite different from
preheating based on parametric resonance.
It turns out that there is very efficient  tachyonic instability that
appears in the  hybrid inflation models. The backreaction of
rapidly generated fluctuations does not allow homogeneous background
oscillations to occur because all energy of the oscillating field is
transferred to the energy of long-wavelength scalar field fluctuations
within a single oscillation!  However, this does not
preclude the subsequent decay of the Higgs and inflaton
inhomogeneities  into
other particles, and thus reheating without parametric resonance.

Let us first consider the background evolution and the results of the naive
perturbative approach to decsribe the quantum fluctuations around the background solutions.
Consider  the simple
potential for the two-field hybrid inflation is
\begin{equation} \label{hyb_eqn}
V(\phi, \sigma) = {\lambda \over 4} (\sigma^2 - v^2)^2 + {g^2 \over
2} \phi^2 \sigma^2  \, ,
\end{equation}
where we used notations  $\Phi_1=\phi$,  $\Phi_2=\sigma$.
Inflation in this model occurs while the homogeneous $\Phi_1$ field
slow rolls from large $\phi$ towards the bifurcation point at
$\phi = {\sqrt{\lambda} \over g} v $ (due to the slight lift of
the potential in $\phi$ direction).  Once $\phi(t)$ crosses the
bifurcation point, the curvature of the $\sigma$ field,
$m^2_{\sigma} \equiv \partial^2 V/\partial \sigma^2$, becomes
negative. This negative curvature results in exponential growth of
$\sigma$ fluctuations. Inflation then ends abruptly in a
``waterfall'' manner.

One reason to be interested in hybrid inflation is that it can be
implemented in supersymmetric theories (\ref{lag}). In particular, for
illustration we will consider preheating in the
 supersymmetric F-term inflation as an example
of a hybrid model.

The simplest F-term hybrid inflation model (without undesirable domaine walls)
is based on a superpotential
with three  left-chiral superfields $\Phi_i=(\Phi_1, \Phi_2, \Phi_3)$
in the Lagrangian (\ref{lag})
\begin{equation}\label{super}
W = {\sqrt{\lambda}\over2}\Phi_1\left(4\Phi_2 \Phi_3 - v^2\right) \, .
\end{equation}
In this case, the spontaneous breaking of the local (global) $U(1)$ symmetry
between the $\Phi_2$ and $\Phi_3$ fields will lead to gauge (global)
string formation.

In global SUSY, using the same notation for superfields and their
complex scalar components, this superpotential contributes
\begin{equation} \label{pot_fterm}
V_{\rm F} = {\lambda\over4} |4\Phi_2\Phi_3 - v^2|^2 +
4\lambda |\Phi_1|^2 \left(|\Phi_3|^2 + |\Phi_2|^2\right) \, .
\end{equation}
to the scalar potential. In general, $\Phi_3$ and $\Phi_3$ could
be (oppositely) charged under a local $U(1)$ symmetry, in which case we
should include a D-term, $V_{\rm D}$,
which we neglect here.

In this model, inflation occurs when chaotic initial conditions lead to
$\langle |\Phi_1 | \rangle \gg v$. When this happens, the  fields $\Phi_2$ and
$\Phi_3$
acquire large effective masses and roll to their local minimum at $\langle
\Phi_2\rangle = \langle\Phi_3\rangle =0$.  In this limit, the
potential~(\ref{pot_fterm}) becomes $V \approx {\lambda v^4 \over 4}$,
which gives rise to a non-vanishing effective cosmological constant.
However, this is a false vacuum state; the true vacuum corresponds to
$\langle\Phi_2\Phi_3\rangle = {v^2 \over 4}$ and $\langle \Phi_1
\rangle = 0$.  The slow-roll potential drives the evolution of
the inflaton towards its true VEV.  When its
magnitude reaches the value $\langle|\Phi_{\rm c}|\rangle = {v\over2}$
spontaneous symmetry breaking occurs.

For further discussion of symmetry breaking in this model, let us
rewrite~(\ref{pot_fterm}) in terms of
polar fields: $\Phi_3=| \Phi_3 | e^{i \theta}$,
${\Phi_2}=|\Phi_2  | e^{i {\bar \theta}}$.  The potential becomes
\begin{equation} \label{potff}
V_{\rm F} = {\lambda \over 4}\left( 16|\Phi_2|^2 |\Phi_3|^2 -
8v^2|\Phi_2||\Phi_3| \cos(\theta+\bar\theta) + v^4\right) +
4\lambda |\Phi_2|^2 \left( |\Phi_2|^2 + |\Phi_3|^2\right) \, .
\end{equation}
At the stage of symmetry breaking, when $\langle | \Phi_2 \Phi_3 | \rangle$
begins to move away from zero, the absolute phase
${\rm Arg}(\Phi_2 \Phi_3 ) = \theta+\bar\theta$ acquires a mass and is
forced to zero.  Note, however, that the potential is independent of the
relative phase, $\theta-\bar\theta$, reflecting the $U(1)$ symmetry.
Thus, in a quasi-homogeneous patch, the $U(1)$
symmetry allows us to choose the relative phase of the $\Phi_2$, $\Phi_3$
 fields
to be zero without any loss of generality.  This choice, combined with the
vanishing of the absolute phase, is equivalent to choosing the two complex
$\Phi_2$, $\Phi_3$  fields to be real. In order to leave
 canonical kinetic terms,
we define $\sigma_\pm \equiv |\Phi_3| \pm |\Phi_2|$.  Furthermore,
as inflation has left the inflaton homogeneous across all the patches, we
may choose it to be real: $\phi \equiv \sqrt2|\Phi_1|$.

In terms of these three real fields, the potential now becomes
\begin{equation}\label{potf}
V_{\rm F} = {\lambda \over 4} \left(\sigma_+^2 - \sigma_-^2 -
v^2\right)^2 + \lambda \phi^2 \left(\sigma_+^2 + \sigma_-^2\right) \ .
\end{equation}
In the symmetric phase, when $\sigma_\pm = 0$, the $\sigma$ fields
have an effective mass-squared: $m_\pm^2(\phi) = \lambda
\left( 2 \phi^2 \mp v^2 \right)$.
We can now see that spontaneous symmetry breaking
 occurs in this model exactly as in the
two field model (\ref{hyb_eqn}).  For
$\phi < \phi_c = {v \over \sqrt{2}}$, the $\sigma_+$ field has a
tachyonic mass that triggers symmetry breaking and the end of inflation.
On the other hand, the $\sigma_-$ field
has always a large and positive effective mass-squared, pinning it
to zero.  Thus, during inflation and at the {\em initial} stages of symmetry
breaking, this model behaves just like the standard
two field hybrid model (\ref{hyb_eqn}).  We have only to apply
the constraint $g^2 = 2\lambda$ and identify the Higgs field
with $\sigma_+$.
The equations for the homogeneous background components
$\phi(t)$ and
$\sigma_\pm(t)$ admits simple solution
\begin{eqnarray}\label{combination}
 \phi(t)+{ 1 \over \sqrt{2}}\sigma_+(t)=\phi_c \ ,  \hspace{2cm}
 \sigma_-(t)=0 \ .
\end{eqnarray}

To study preheating in the F-term inflation, we have to
 analyse evolution of the  vacuum fluctuations.  Consider  vacuum
fluctuations   in the inflaton sector $\Phi_i$ of the theory (\ref{potff}).
Usual description in terms of  a homogeneous background plus small
fluctuations gives us equations for fluctuations around the background
solution  (\ref{combination}). We  define the
variances of fields:
\begin{equation}
\left< \,  \vert \sigma_\pm - \left< \sigma_\pm \right> \vert^2 \,
\right>_{\rm ren} = \int{d^3k \over {(2 \pi)^3}} \, \left[ \,
\vert \delta\sigma_{k\pm}(t) \vert^2 - \vert \delta\sigma_{k\pm}(0) \vert^2
\, \right] \equiv \int{{{dk} \over k}} \, {\mathcal P}_{\pm}(k,t)
\end{equation}
and similar for $\phi$ field.
Here $ {\mathcal P}_{\pm,\phi}(k,t)$ are the spectra of the fluctuations.

\begin{figure}[t]
\centering \leavevmode \epsfxsize=7.5cm    %7.2cm
\epsfbox{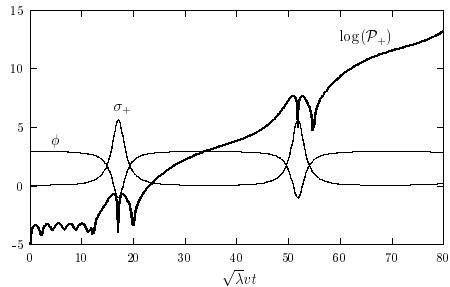}\\
\caption{Evolution of the background fields
 $\sigma_+(t)$ and $\phi(t)$ after symmetry
breaking and the $\log$ of ${\mathcal P}_+(k)$ for the mode with the momentum
$k = 0.2 \sqrt{\lambda} v$.
%The vertical scale is arbitrary for the field amplitudes.
} \label{bkgd}
\end{figure}

\begin{figure}[b]
\centering \leavevmode \epsfxsize=7.5cm
\epsfbox{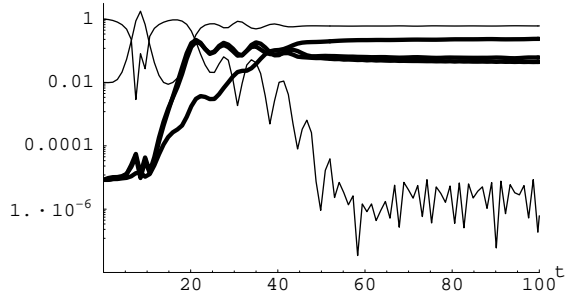}\\
\caption{Means and variances in units of $\phi_c$.
 The squared means $\langle\phi\rangle^2$
and $\langle \sigma_+ \rangle^2$ are ordinary solid lines while the field
variances  are thick lines. The mean of
$\phi $ starts at $\phi_c$, oscillates once, and then decays.
 The mean of $\sigma_+$ grows in antiphase
to $\phi$ and freeses at $\phi_c$.} \label{potmeansvars}
\end{figure}

 Numerical solutions of the equations  for
the background fields and
  for the time
evolution of the linear  fluctuations of $\sigma_+$
for the mode $k$ where their spectrum is maximum
 are plotted as
the bold line at the Figure  \ref{bkgd}.
Notice  an enormous exponential
growth of the fluctuations within a single background
oscillation. Indeed, the amplitude  ${\mathcal P}_+(k)$ increases by
factor $10^{10}$!
 There are two factors which contributes to such a
strong  instability of fluctuations in the model. First,
oscillating background fields are crossing the region with significant
negative curvature of the effective potential, which
results in  tachyonic instability.
  Second, this region
turn out to be a turning point for the background oscillations, where
the fields spend significant portion of the oscillation.  As a result,
tachyonic instability is lasting long enough to make  the
backreaction of the fluctuations to be significant already within
single background oscillation.  The regime of background oscillations
 will even not be settled.
Therefore practically from the beginning we have to
use the lattice simulations to study  nonlinear
dynamics of the fields.

The results of full nonlinear lattice simulations in the model
derived in \cite{GBFKLT}
are plotted in Figure \ref{potmeansvars}.
The simulations showed that the homogeneous fields
${\phi}$ and ${\sigma}_+$ initially followed the classical trajectory
(\ref{combination}) but, within one oscillation of the inflaton field,
fluctuations grew too large to speak meaningfully of the fields as
homogeneous oscillators. These fluctuations grew in such a way
that ${\sigma_+}={{\sigma_-}}^*$ almost exactly
throughout the simulation. In other words
Re $\delta{\sigma}_+$ and Im $ \delta {\sigma}_-$ were
excited while Im $ \delta {\sigma}_+$ and
Re $\delta {\sigma}_-$ were not. Because of this we only plot
the fields ${\phi}$ and ${\sigma_+}$.

In \cite{GBFKLT,FKL} we develop a general theory of tachyonic
 preheating, which occurs due to
tachyonic instability in the theories with spontaneous symmetry breaking.
Our approach combines analytical estimates with lattice simulations
taking into account all backreaction effects. The process of spontaneous
symmetry breaking involves transfer of the potential energy into the
energy of fluctuations produced due to the tachyonic instability. We show
that this process is extremely efficient and requires just a single
oscillation of the scalar field falling from the top of the effective
potential. In what follows I will illustrate tachyonic preheating
using the model (\ref{super}). 

\section{Spontaneous Symmetry Breaking and Tachyonic Preheating}

To understand physics of tachyonic preheating, we have to go back to the basics of the
spontaneous symmetry breaking \cite{FKL}.
The simplest model of spontaneous symmetry breaking is based on the
theory with the  effective potential
$V_{\rm F} = {\lambda \over 4}v^4+
{\lambda \over 4} \sigma_+^4 -{\lambda \over 4} \sigma_+^3  + \lambda \sigma_+^2$,
dominated by the quadratic term near the top of the potential.
The development of tachyonic instability in this model depends on
the initial conditions. We will assume that initially the symmetry
is completely restored.
Initially  scalar field fluctuations in this  model
 in the symmetric phase $\phi=0$  are the same as for a massless field,
$\phi_k ={1 \over \sqrt{2 k}}e^{-ikt +i{\vec k \vec
x}}$. Then at $t = 0$ we `turn on' the term $-m^2\phi^2/2$
corresponding to the negative mass squared $-m^2$. The modes with
$k = |{\vec k}|  < m$ grow  exponentially. 
Initial dispersion of all growing fluctuations with $k < m$ was given by $
\langle \delta\phi^2 \rangle
 =  \int\limits_0^{m}  {  dk^2  \over 8\pi^2 } = {m^2\over 8 \pi^2}$
and the average initial amplitude of all fluctuations with $k < m$ was given by 
$ \delta\phi
 =  {m \over 2 \pi }$
The  dispersion of the growing modes at  $t > 0$ is growing exponentially.
This means that the average amplitude $\delta\phi(k)$ of quantum fluctuations with momenta 
  $\sim k$ initially was $\delta\phi(k) \sim k/2\pi$, and then it started
 growing as $e^{t\sqrt{m^2- k^2}}$.
The tachyonic growth of all fluctuations with $k <
m$ continues until $\sqrt{\langle
\delta\phi^2 \rangle}$ reaches the value $\sim v/2$, since at
$\phi \sim v/\sqrt 3$ the curvature of the effective potential
vanishes and instead of tachyonic growth one has the usual
oscillations of all the modes. This happens
within a time $\Delta t 
\sim {1\over 2 m} \ln{C\over \lambda}$, where $C \sim 10^2$.

The process of symmetry breaking  will occur in a somewhat different way in theories
where the curvature of the effective potential near its maximum depends on
$\phi$. Consider for example $V =  -{\lambda\over 4} \phi^4$ near the top of the potential.
In this case there is an instanton solution corresponding to the
 tunneling  from $\phi=0$ to nonzero value \cite{Fub,Linde1}.

Let us now consider the tachyonic instability for the theory 
 (\ref{super}). 
The scalar fields potential (\ref{potf}) for the solution (\ref{combination})
in  the fields space is reduced to the potential 
\begin{equation}\label{potfn}
V_{\rm F} = {\lambda \over 4}v^4-\lambda \sigma_+^3+{3 \over 4}\lambda \sigma_+^4
\end{equation}
Let us simplify notation $\sigma_+ \to \phi$ and make rescaling $\lambda \to \lambda/3$.
Then the potential (\ref{potfn}) once again is reduced to the form 
\begin{equation}
V= -{\lambda\over 3} v\phi^3+{\lambda\over 4}\phi^4+{\lambda\over 12} v^4 \
.
\label{cub}
\end{equation}
Thus the  theory (\ref{cub}) is 
 a prototype of the theory  (\ref{super}), which initially looked much more complicated.

Development of instability in the  theory (\ref{cub}) presents us with a new challenge.
The curvature of the effective potential at $\phi = 0$ in this theory
vanishes, which means that, unlike in the theory $-m^2\phi^2$,
infinitesimally small perturbations in this theory do not grow. On the
other hand, in this theory, unlike in the theory $-\lambda\phi^4$,
 there are no instantons which would describe tunneling
from $\phi = 0$. Thus, in the theory $-{\lambda } v\phi^3$, which occupies
an  intermediate position between $-m^2\phi^2$ and $-\lambda\phi^4$, both
mechanisms which could lead to the development of instability do not work.
Does it mean that the state $\phi = 0$ in this theory is, in fact, stable?

The answer to this question is no, the state $\phi = 0$ in the theory
$-{\lambda } v\phi^3$ is unstable. Indeed, even though $\langle\phi\rangle$
initially is zero, long wavelength fluctuations  of the field $\phi$ are
present, and they may play the same role as the homogeneous field $\phi$ in
triggering the instability.

The  scalar field
fluctuations with momentum $k < k_0$  have initial amplitude
$\langle \delta\phi^2 \rangle \sim {k_0^2\over 8\pi^2}$. Thus the short
wavelength fluctuations with momenta $k > k_0$ live on top of the long
wavelength field with an average amplitude $\delta\phi_{\rm
rms}(k_0) \sim \sqrt {\langle \delta\phi^2 \rangle} \sim {k_0 \over 2 \sqrt
2 \pi }$.

The curvature of the effective potential $V'' = |m^2_{\rm eff}|$ at $\phi
\sim \delta\phi_{\rm
rms}(k_0)$ in the theory (\ref{cub}) is given by $-2\lambda v
\delta\phi_{\rm
rms}(k_0) \sim - \lambda v  {k_0 \over \sqrt 2 \pi }$. Consider the
fluctuations with momentum $k$  somewhat greater than $k_0$, so that the
amplitude of the long wavelength field $\delta\phi$ does not change
significantly on a scale $k^{-1}$. Short wavelength fluctuations with $k =
C k_0$ with $C$ somewhat greater than $1$ will grow on top of the field
$\phi \sim \delta\phi_{\rm
rms}(k_0)$ if $k^2 < |m^2_{\rm eff}|   \sim {\lambda v  k_0
\over \sqrt 2\pi} $.

Taking for definiteness $C \sim \sqrt 2$, one may argue that
fluctuations with $k < {\lambda v \over 2\pi}$ may enter a
self-sustained regime of tachyonic growth. Small fluctuations rapidly grow
large, which  justifies semi-classical methods used for the description of
this process. The average initial amplitude of the growing tachyonic
fluctuations with
momenta smaller than ${\lambda v \over 2\pi}$ is
\begin{equation}\label{typical}
 \delta\phi_{\rm rms} \sim {\lambda v   \over 4\pi^2}.
\end{equation}
These fluctuations grow until the amplitude of $\delta\phi$
becomes comparable to $2v/3$, and the effective tachyonic mass
vanishes. At that moment the field can be represented as a
collection of waves with dispersion $\sqrt{\langle
\delta\phi^2\rangle} \sim v$, corresponding to coherent states of
scalar particles with occupation numbers $n_k \sim
\left({4\pi^2\over \lambda}\right)^2 \gg 1.$ A more accurate investigation
shows that the initial value of the field is few times greater than
$\sqrt{\langle
\delta\phi^2\rangle} \sim v$, and therefore the occupation
numbers will be somewhat smaller,
\begin{equation}\label{occcub}
n_k \sim O(10)\,  \lambda^{-2} \ .
\end{equation}

Because of the nonlinear dependence of the tachyonic mass on
$\phi$, a detailed description of this process is more involved
than in the quadratic theory. Indeed, even though the typical
amplitude of the growing fluctuations is given by (\ref{typical}),
the speed of the growth of the fluctuations increases considerably
if the initial amplitude is somewhat bigger than (\ref{typical}).
Thus even though the fluctuations with amplitude a few times
greater than (\ref{typical}) are exponentially suppressed, they
grow faster and may therefore have greater impact on the process
than the fluctuations with  amplitude (\ref{typical}).

\begin{figure}[t]
%\begin{figure}[Fig001]
%\centering
\leavevmode\epsfysize=6.5cm \epsfbox{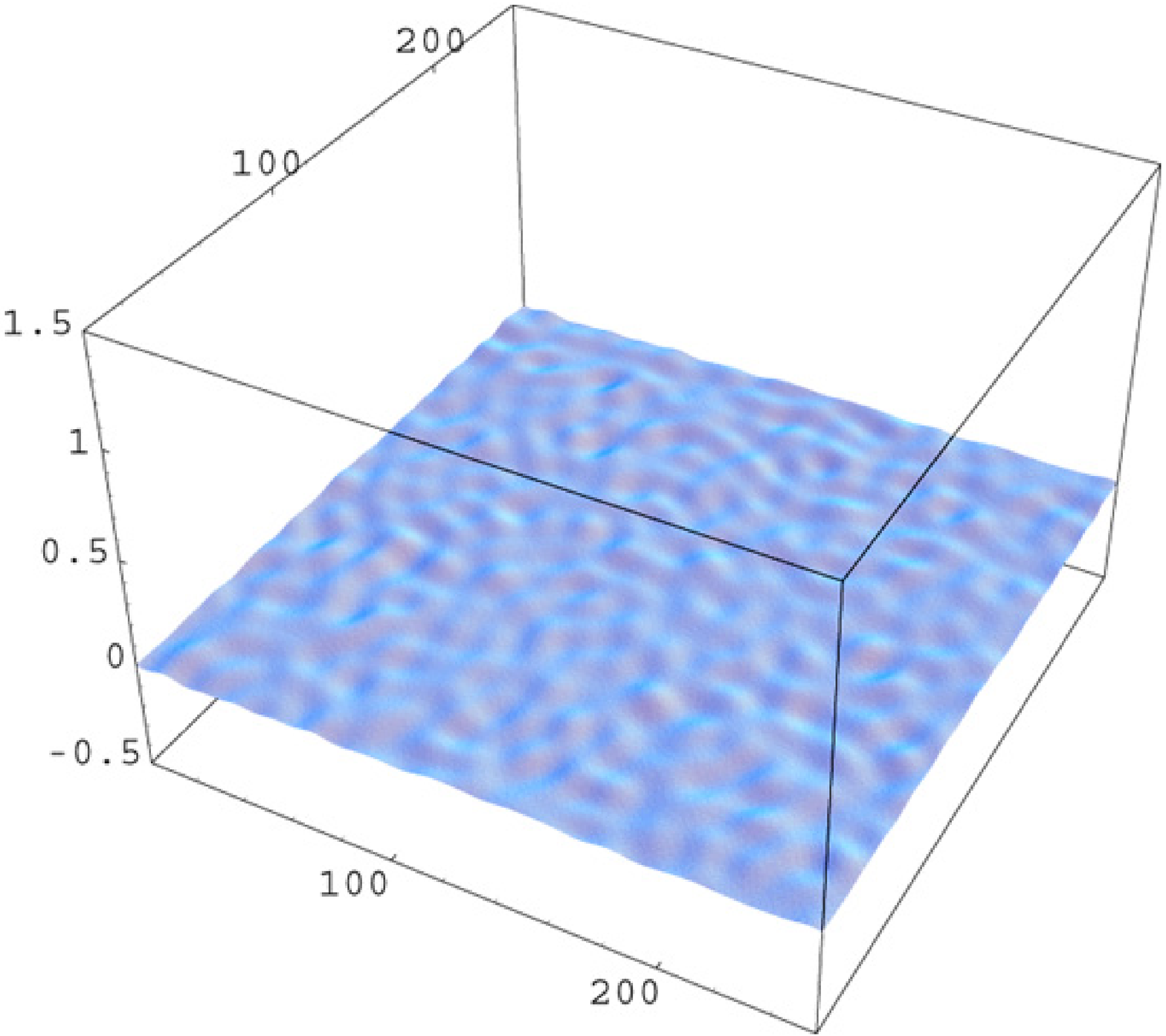}\hskip 0.1 cm
\leavevmode\epsfysize=6.5cm  \epsfbox{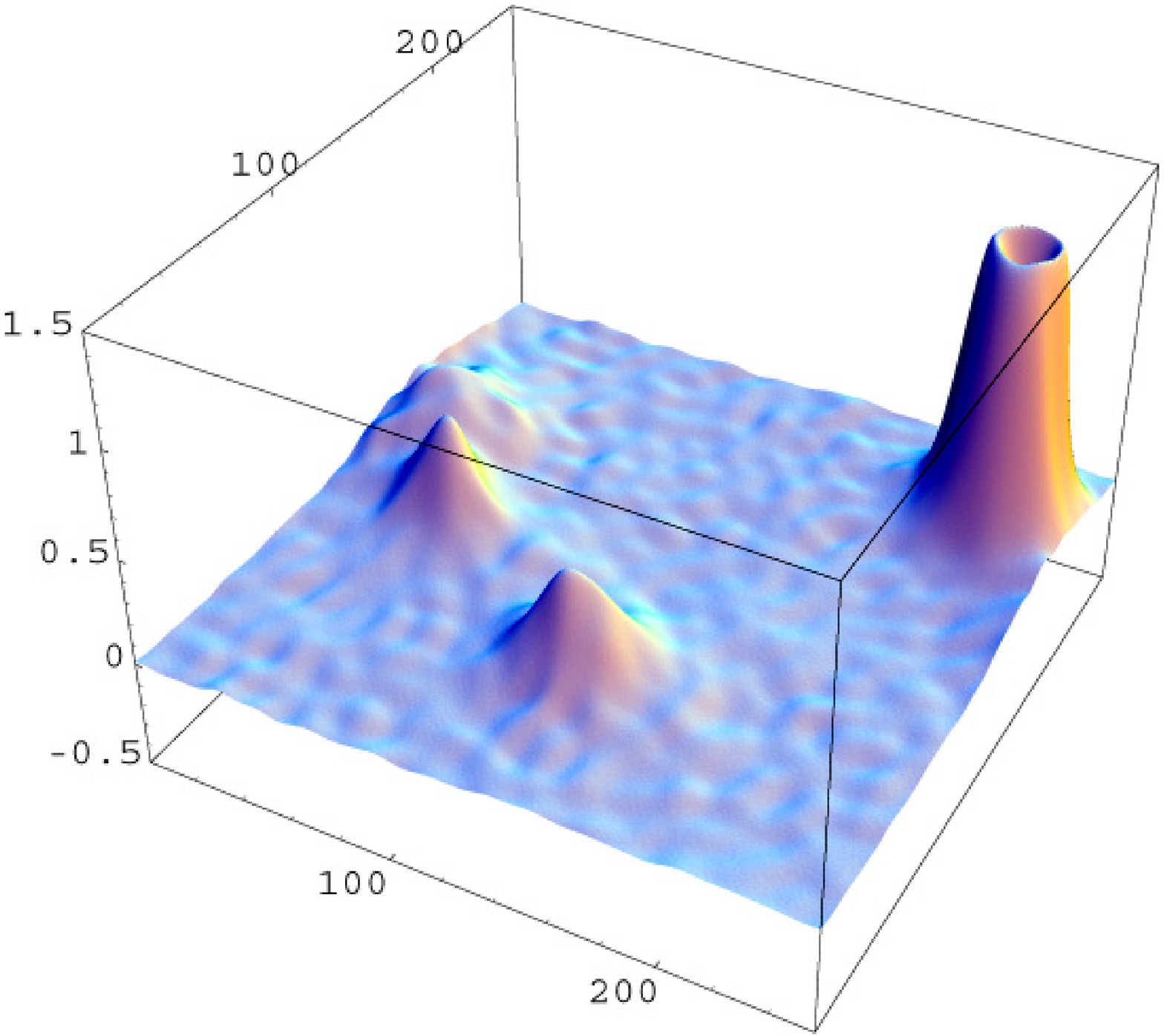}
%\picdir{pdf_lmb=1e-8.eps
\end{figure}
%\vskip -0.8cm

\begin{figure}[b]
%\begin{figure}[Fig001]
%\centering
\leavevmode\epsfysize=6.5cm \epsfbox{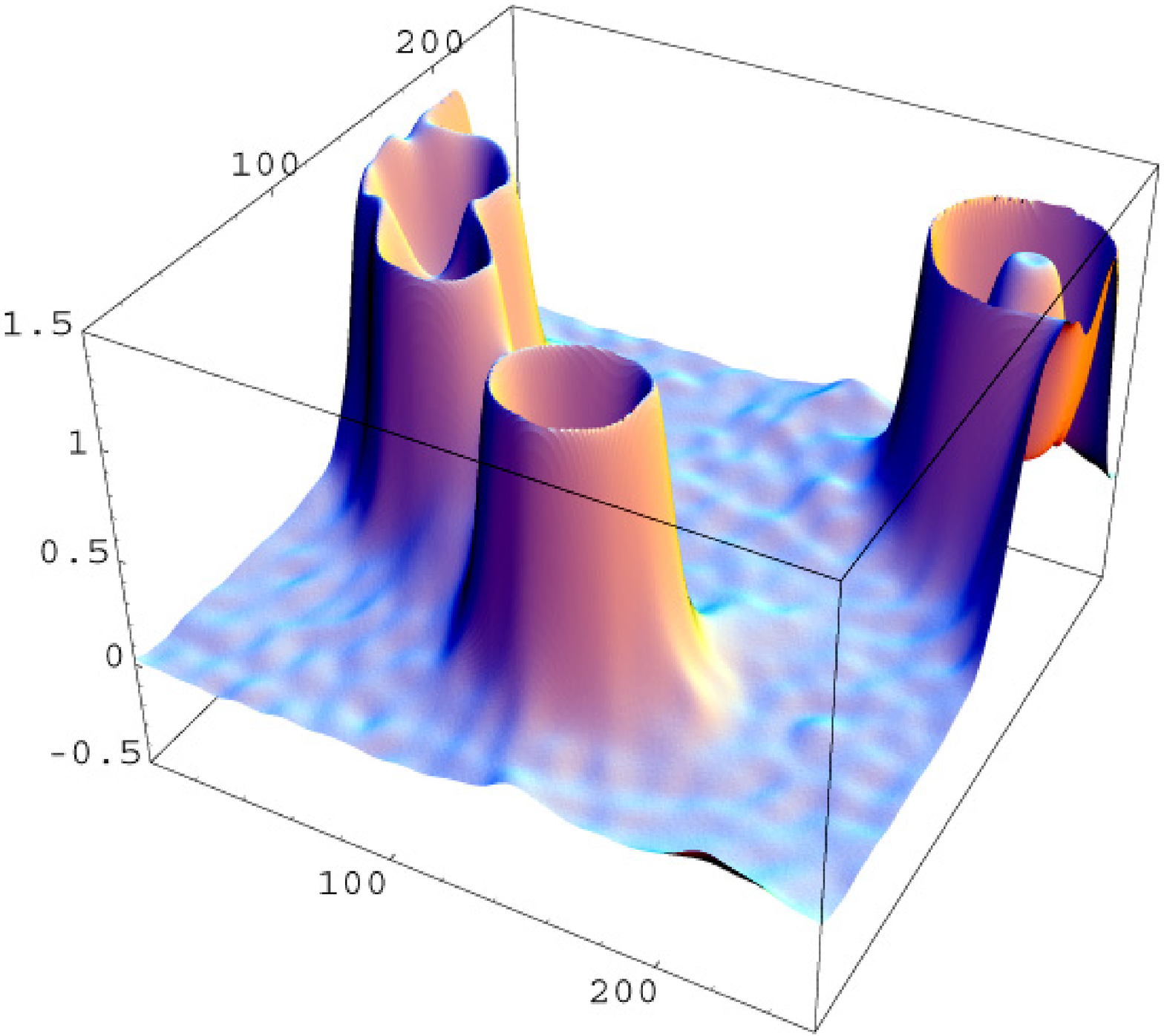}\hskip 0.1 cm
\leavevmode\epsfysize=6.5cm  \epsfbox{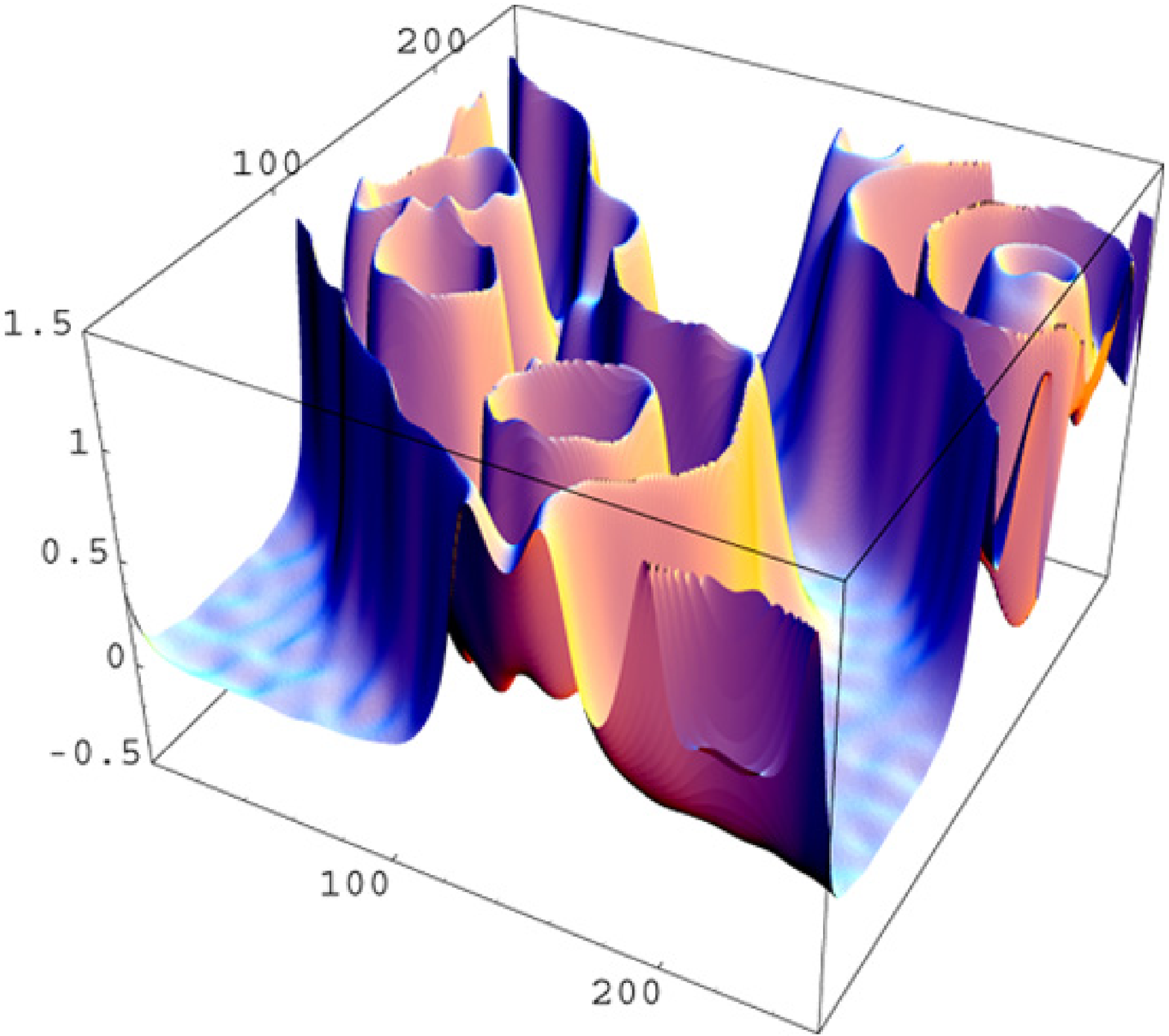}
%\picdir{pdf_lmb=1e-8.eps
\end{figure}

%\vskip -0.8cm
\begin{figure}[t]
%\begin{figure}[Fig001]
%\centering
\leavevmode\epsfysize=6.5cm \epsfbox{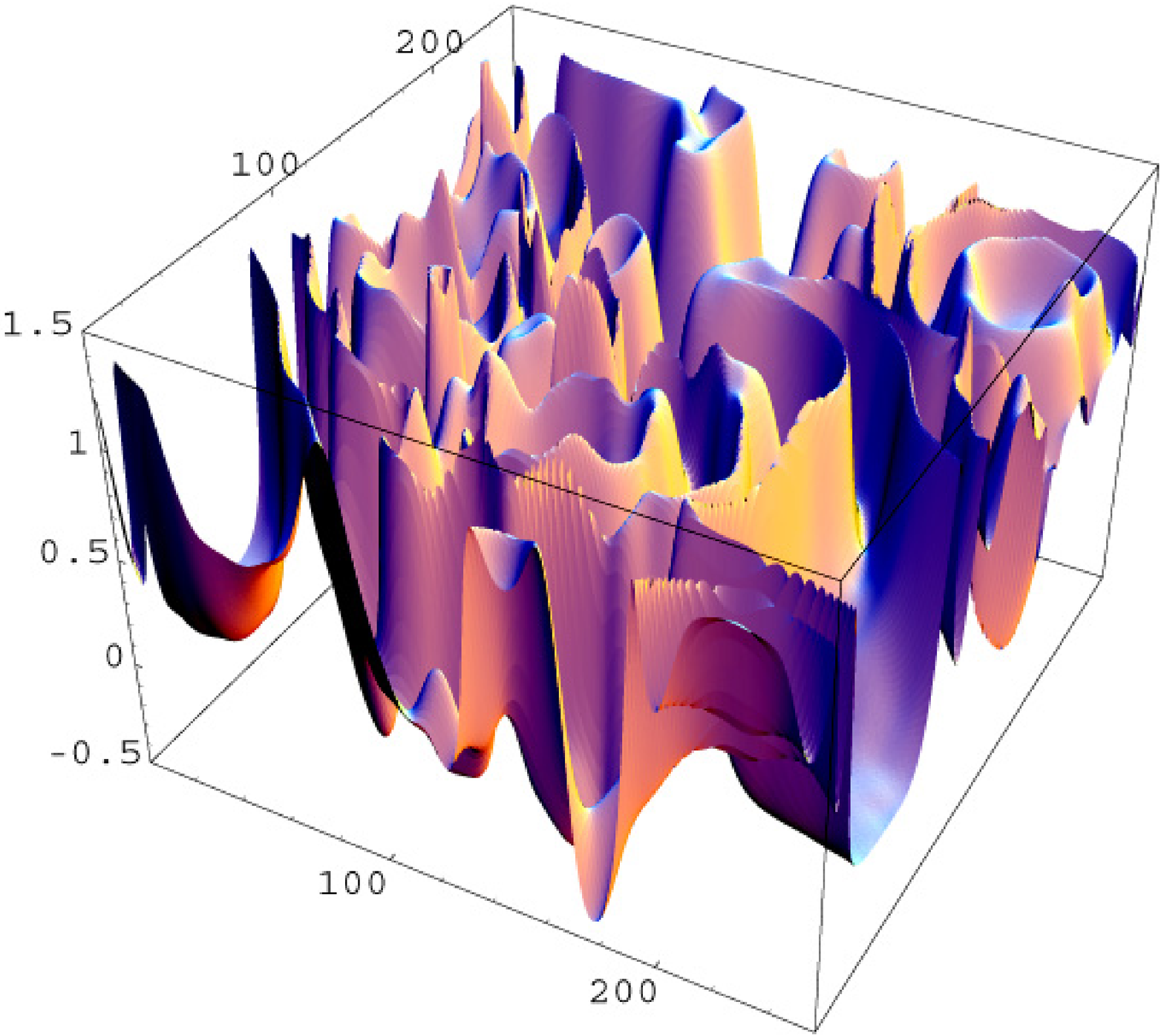}\hskip 0.1 cm
\leavevmode\epsfysize=6.5cm  \epsfbox{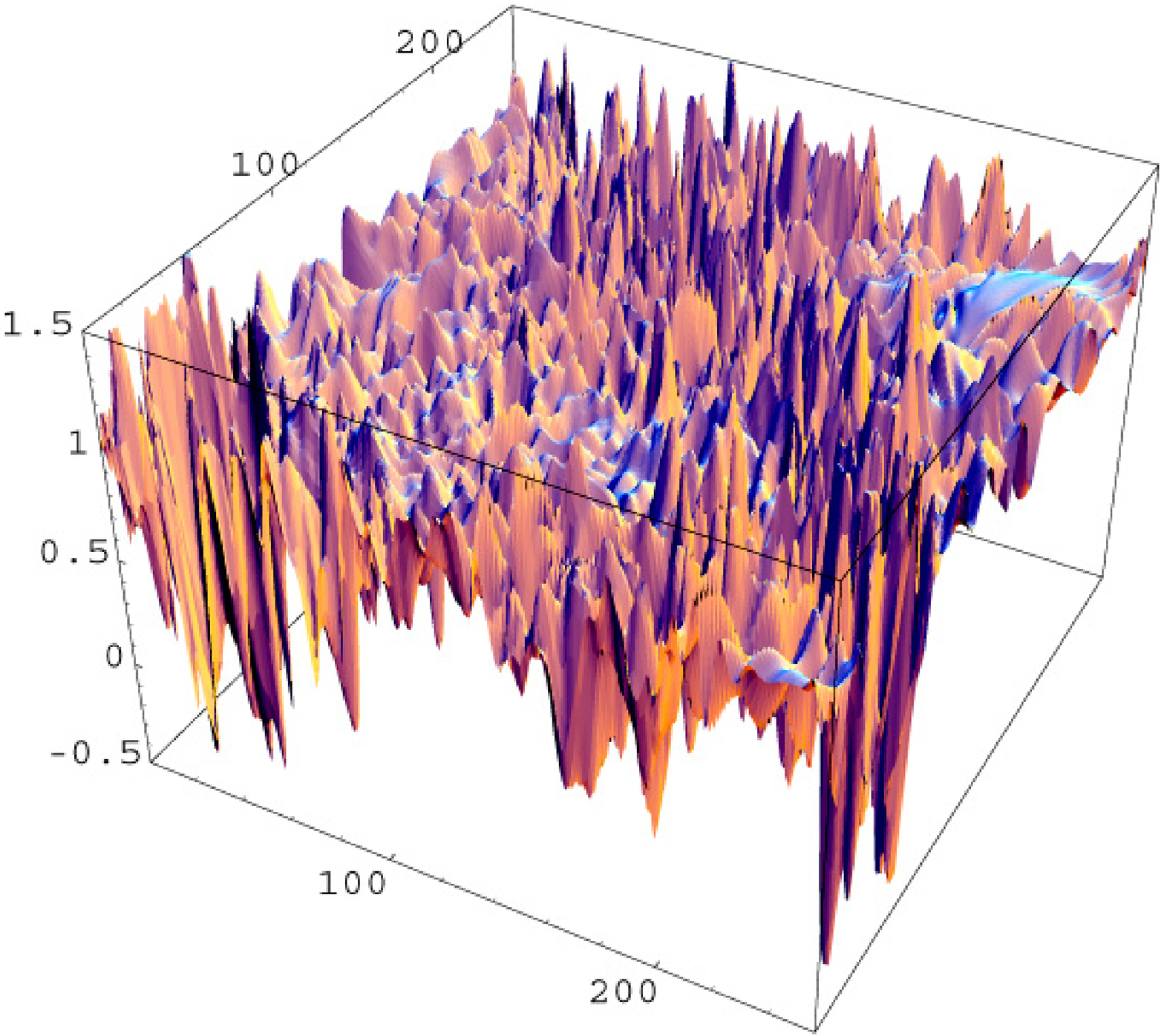}
%\picdir{pdf_lmb=1e-8.eps

\
%\caption[Fig001]
\caption{\label{2gif} Three panels above show  growth of quantum fluctuations of the field
$\phi$ in the theory  $V= -{\lambda\over 3} v\phi^3+{\lambda\over 4}\phi^4$
(\ref{cub}) looks like bubble formation. Preheating occurs due to a
combined effect of bubble production, tachyonic instability and bubble wall
collisions. }
\end{figure}

Low
probability fluctuations with $ \delta \phi \gg \delta\phi_{\rm
rms}$  correspond to peaks of the initial Gaussian distribution of
the fluctuations of the field $\phi$. Such peaks tend to be
spherically symmetric \cite{BBKS}.   As a result, the whole
process looks not like a uniform growth of all modes, but more
like  bubble production (even though there are no instantons in
this model).

To study this issue in a more detailed way, one may use a stochastic
approach to tunneling and bubble formation developed in \cite{Linde2}. The
main idea of this approach can be explained as follows. Tunneling can be
represented as a result of accumulation of quantum fluctuations with the
amplitude greatly exceeding their usual value determined by uncertainty
principle. This happens when the long wavelength  quantum fluctuations
responsible for the tunneling correspond to bosonic excitations with large
occupation numbers. In such cases one can treat these fluctuations as
classical fields  experiencing Brownian motion due to their interaction
with the short wavelength quantum fluctuations.

Suppose that the large fluctuations of the scalar field responsible for
reheating in the model (\ref{cub}) initially look like spherically
symmetric bubbles (which is the case if the probability of such
fluctuations is strongly suppressed, see above). Equation of motion for a
bubble of a scalar field $\phi(r)$ in  Minkowski space is
\begin{equation}\label{10}
\ddot\phi = \phi''  +  2  \phi' r^{-1}  - V'(\phi) \  .
\end{equation}
Here $r$ is a distance from the center of the bubble, $\phi' =
{\partial\phi\over \partial r}$.
At the moment of its formation, the bubble wall does not move, $\dot\phi =
0$, $\ddot\phi = 0$ (critical bubble). Then it gradually starts
growing,  $\ddot\phi > 0$, which requires that
\begin{equation}\label{11}
 | \phi'' +  2 \phi' r^{-1}| < - V'(\phi) \ .
\end{equation}
A bubble of a classical field is formed only if  it
contains a sufficiently large field $\phi$, and if the bubble itself is
sufficiently large. If the size of the
bubble is too small, the gradient terms are greater than the term
$|V'(\phi)|$, and
the field $\phi$ inside the bubble does not grow.

At small $r$ the shape of the bubble can be approximated by $\phi = \phi(0)
-\alpha r^2/2$. In this approximation, the bubble has a typical size $r_0
\sim \sqrt{2\phi(0)\over \alpha}$, and  $\phi' r^{-1} = \phi'' = -\alpha $.
Therefore at the moment of the bubble formation, when $\ddot \phi = 0$, one
has
\begin{equation}\label{11a}
 \phi''   = V'(\phi(0))/3 \ .
\end{equation}
Replacing $\phi''$ by $k_0^2 \phi(0)$ one finds that the bubble can be
considered a result of overlapping of quantum fluctuations with typical
momenta $k < k_0 \sim r_0^{-1}$, where
\begin{equation}\label{11aa}
 k^{2}_0 = C^2 {V'(\phi(0))\over 3\phi(0) } \ .
\end{equation}
Here $C = O(1)$ is some numerical factor reflecting uncertainty in our
estimate of $k_0$.

Let us estimate the probability of an event when vacuum fluctuations
occasionally build up a configuration of the field satisfying this
condition.
In order to do it one should remember that the dispersion of quantum
fluctuations of the  field $\phi$ with $k < k_0$ is given by $\langle
\delta\phi^2 \rangle \sim {k_0^2\over 8\pi^2}$.
This gives
\begin{equation}\label{14}
\langle \phi^2 \rangle_{k<k_0} \sim {k_0^2\over 8\pi^2} =  C^2
{V'(\phi(0))\over 24 \pi^2 \phi(0) } \ .
\end{equation}

This is an estimate of the dispersion of perturbations which may sum up
to produce a bubble of the field $\phi$ that satisfies the condition
(\ref{11}).  Of course,
this estimate is rather crude.  But
let us  nevertheless use eq. (\ref{14}) to evaluate the probability that
these
fluctuations build up a bubble of a radius  $r >
k^{-1}_0$ containing the field $\phi$ at its center. Assuming, in the first
approximation, that the probability distribution is gaussian, one finds:
\begin{equation}\label{15}
P(\phi) \sim \exp\left(-{\phi^2\over
2\langle \phi^2 \rangle_{k<k_0}}\right) =
\exp \left(-{12\pi^2 \phi^3\over C^2 V'(\phi)}\right) \ .
\end{equation}

The general formula (\ref{15}), being applied to
the theory $-\lambda\phi^4/4$
 to within a factor of
$C \approx 2$  coincides with the Euclidean action
for the instanton in this theory.
Taking into account  the very rough method we used to estimate $k_0$ and
calculate the dispersion of the perturbations responsible for tunneling,
the coincidence is rather impressive. As it was shown in
\cite{StarStoch,Linde2}, in application to the tunneling during inflation
in the potentials with $V'' \ll H^2$ this approach gives exactly the same
answer as the Euclidean approach.
ost importantly, this methods allows   to investigate tunneling and
development of instability in
the theories where the instanton solutions do not
exist \cite{Linde2}. In particular, for the tunneling in the theory
$-\lambda v \phi^3/3$ one finds
\begin{equation}\label{16}
P(\phi) \sim  (\lambda v \phi)^2  \exp \left(-{12\pi^2 \phi \over C^2 \lambda
v}\right) \ .
\end{equation}
We included here the subexponential factor $O(k_0^4) \sim (\lambda v
\phi)^2$, which is necessary to describe the probability of tunneling per
unit time per unit volume.

This means that the tunneling is not suppressed for $\phi \sim {C^2 \lambda
v\over 12\pi^2}$. This result is in agreement with our previous estimate
(\ref{typical}). Now let us take into account that the total time of the
development of instability is a sum of the time of tunneling plus the time
necessary for rolling of the field down. One can show that the time of
rolling down is inversely proportional to $m(\phi) \sim \sqrt{\lambda v
\phi}$, i.e. it decreases at large $\phi$. Also, the subexponential factor
$(\lambda v \phi)^2$ grows at large $\phi$, which makes tunneling to large
$\phi$ faster.  Consequently, as we already discussed above, the main
contribution to the development of instability is given by the fluctuations
with $\phi >   \sim {C^2 \lambda v\over 12\pi^2}$. Exponential suppression
of the probability of such fluctuations leads to their approximate
spherical symmetry.

The results of our lattice simulations for this model \cite{FKL}
are shown in three panel of  Fig. \ref{2gif}.  In this model bubbles form quickly
enough that we were able to start with quantum fluctuations centered at
$<\phi>=0$ and allow the bubbles to form automatically. The bubbles (high
peaks of the
field distribution) grow, change shape, and interact with each
other, rapidly  dissipating the vacuum energy $V(0)$.

\section{Development of Equilibrium after Preheating}

   The character of preheating may vary from model
to model, e.g. parametric excitation in chaotic inflation
\cite{KLS} and tachyonic  preheating  in hybrid
inflation \cite{GBFKLT}, but its distinct feature remains the same:
rapid amplification of one or more bosonic fields to exponentially
large occupation numbers. This amplification is eventually shut
down by backreaction of the produced fluctuations. The end result
of the process is a turbulent medium of coupled, inhomogeneous,
classical waves far from equilibrium.
Despite the development of our understanding of preheating after
inflation, the transition from this stage to a  hot Friedmann
universe in thermal equilibrium has remained relatively poorly
understood.
The details of this thermalization stage
depend on the constituents of the fundamental Lagrangian
(\ref{lag})  and their
couplings, so at first glance it would seem that a description of
this process would have to be strongly model-dependent.
Recently we performed  a fully nonlinear study of the development of
equilibrium after preheating \cite{FK}.
We have
performed lattice simulations of the evolution of interacting
scalar fields during and after preheating for a variety of
inflationary models.
 We have
found, however, that many features of this stage seem to hold
generically across a wide spectrum of models.
Indeed, at the end of preheating and beginning of the turbulent
stage $t_*$, the fields are out of equilibrium. We
have examined many models and found that at $t_*$ there is not
much trace of the linear stage of preheating and conditions at
$t_*$ are not qualitatively sensitive to the details of inflation.
We therefore expect that this second, highly nonlinear, turbulent
stage of preheating may exhibit some universal, model-independent
features. Although a realistic model would include one or more Higgs-Yang-Mills
sectors, we treat the simpler case of interacting scalars.

   We have numerically
investigated the processes of preheating and thermalization in a
variety of models and determined a set of rules that seem to hold
generically. These rules can be formulated as follows
(in this section we use notations $\phi=\Phi_1$ for the inflaton field
and $\chi$, $\sigma$ for other scalars $\Phi_i$)

\bigskip
\noindent {\it 1.  In many, if not all viable models of inflation there
exists a mechanism for exponentially amplifying fluctuations of at
least one field $\chi$. These mechanisms tend to excite
long-wavelength excitations, giving rise to a highly infrared
spectrum.}

The mechanism of parametric resonance in single-field models of
inflation has been studied for a number of years.
This effect is quite robust. Adding
additional fields (e.g.  $\sigma$ fields) or self-couplings
(e.g. $\chi^4$) has little or no effect on the resonant period.
Moreover, in many hybrid models a similar effect occurs due to
tachyonic  instability. The qualitative features of the fields
arising from these processes seem to be largely independent of the
details of inflation or the mechanisms used to produce the fields.

\bigskip
\noindent {\it 2.  Exciting one field $\chi$ is sufficient to
rapidly drag all other  light fields with which $\chi$
interacts into a similarly excited state.}

We have seen this effect when multiple fields are coupled directly to
$\chi$ and when chains of fields are coupled indirectly to $\chi$. All
it takes is one field being excited to rapidly amplify an entire
sector of interacting fields. These second generation amplified fields
will inherit the basic features of the $\chi$ field, i.e. they will
have spectra with more energy in the infrared than would be expected
for a thermal distribution.

\bigskip
\noindent {\it 3. The excited fields will be grouped into subsets with
identical characteristics (spectra, occupation numbers, effective
temperatures) depending on the coupling strengths.}

We have seen this effect in a variety of models. For example in
the models (\ref{nfldlambda}) which we are going to consider
  the $\chi$ and
$\sigma$ fields formed such a group. In general, fields that are
interacting in a group such as this will thermalize much more
quickly than other fields, presumably because they have more
potential to interact and scatter particles into high momentum
states.

\bigskip
\noindent {\it 4. Once the fields are amplified, they will approach
thermal equilibrium by scattering energy into higher momentum modes.}

This process of thermalization involves a slow redistribution of
the particle occupation number as low momentum particles are
scattered and combined into higher momentum modes. The result of
this scattering is to decrease the tilt of the infrared portion of
the spectrum and increase the ultraviolet cutoff of the spectrum.
Within each field group the evolution proceeds identically for all
fields, but different groups can thermalize at very different
rates.

Here we will illustrate these results with
a simple  chaotic inflation model with a
quartic inflaton potential. The inflaton $\phi$ has a four-legs
coupling to another scalar field $\chi$, which in turn can couple
to two other scalars $\sigma_1$ and   $\sigma_2$. The potential for this
model is
\begin{equation}\label{nfldlambda}
V = {1 \over 4} \lambda \phi^4 + {1 \over 2} g^2 \phi^2 \chi^2 +
{1 \over 2} h_1^2 \chi^2 \sigma_1^2+{1 \over 2} h_2^2 \chi^2 \sigma_2^2
\end{equation}
Preheating in this theory in the absence of the $\sigma_i$ fields
is well studied.  For nonsmall $g^2 \over \lambda$ the field
$\chi$ will experience parametric amplification, rapidly rising to
exponentially large occupation numbers. In the absence of the
$\chi$ field (or for sufficiently small $g$) $\phi$ will be
resonantly amplified through its own self-interaction, but this
self-amplification is much less efficient than the two-field
interaction. The results shown here are for $\lambda = 9 \times
10^{-14}$ (for CMB anisotropy normalization) and $g^2 = 200 \lambda$. When
we add a third field we use $h_1^2 = 100 g^2$ and when we add a
fourth field we use $h_2^2 = 200 g^2$.

One of the most interesting variable to calculate is the
(comoving) number density of particles of the fields $n(t)$
and their  occupation number $n_k$.
The evolution
of the total number density of all particles $n_{tot}$ is an indication of the
physical processes taking place. In the weak interaction limit the
scattering of classical waves via the interaction  ${1 \over
2} g^2 \phi^2 \chi^2$ can be treated using a perturbation
expansion with respect to $g^2$.
The leading four-legs diagrams
for this interaction corresponds to a two-particle collision
$(\phi \chi \rightarrow \phi \chi)$, which conserves $n_{tot}$.
The regime where such interactions dominate corresponds to ``weak
turbulence'' in the terminology of the theory of wave turbulence.
 If we see $n_{tot}$ conserved it will be an
indication that these two-particle collisions constitute the
dominant interaction. Conversely, violation of $n_{tot}(t)=const$
will indicate the presence of strong turbulence, i.e. the
importance of many-particle collisions. Such higher order
interactions may be significant despite the smallness of the
coupling parameter $g^2$ (and others) because of the large
occupation numbers $n_k$. Later, when these occupation numbers are
reduced by rescattering, the two-particle collision should become
dominant and $n_{tot}$ should be conserved.
For a bosonic field in thermal equilibrium with a temperature $T$
and a chemical potential $\mu$ the spectrum of occupation numbers
in the limit of classical waves
is given by
\begin{equation}\label{wave}
n_k \approx {T \over {\omega_k -\mu}} \, .
\end{equation}

\begin{figure}[t]
\centering \leavevmode \epsfxsize=8.0cm
%\leavevmode\epsfxsize=.7\columnwidth \epsfbox{n4fldl.eps}
\epsfbox{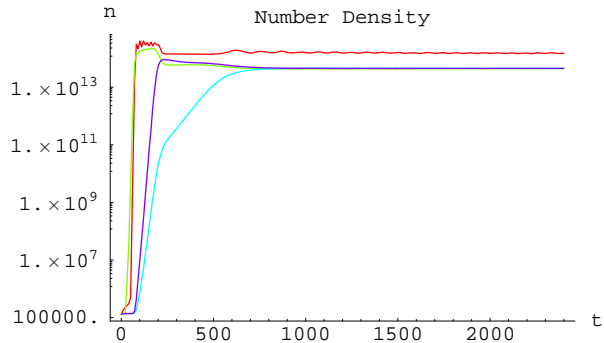}\\
\caption{ Time  evolution of number density of particles in the model
(\ref{nfldlambda}). The curves represent $n_{\phi}$,  $n_{\chi}$,
 $n_{\sigma_1}$, $n_{\sigma_2}$ from top to bottom. Unit of
(conformal) time is  $a \cdot 10^{-36}$ sec.} \label{n4fldl}
\end{figure}

Figure \ref{n4fldl} shows an exponential increase
of  $n(t)$ during preheating, followed by a gradual decrease
 that asymptotically slows down.
This exponential increase is a consequence of
explosive particle production due to parametric resonance.
After preheating the
fields enter a turbulent regime.  In our simulations we see
$n(t)$ decreasing during this stage.
 This decrease is a consequence of the
many-particle interactions beyond the four legs rescattering.

 An important point is that
the interaction of $\chi$
and $\sigma_i$ does not affect the preheating of $\chi_i$, but does
drag $\sigma_i$ exponentially quickly into an excited state.
 The fields $\sigma_i$ are exponentially amplified not by
parametric resonance, but by their stimulated interactions with the
amplified $\chi$ field. Unlike amplification by preheating, this
direct decay nearly conserves particle number, with the result
that $n_\chi$ decreases as $\sigma_i$ grow.

 Interacting waves of scalar fields constitute a dynamical system.
 Dynamical chaos is one of the features of wave
turbulence. In \cite{FK}  we address the question  how and
when the onset of chaos takes place after preheating.
To investigate the
onset of chaos  we have to follow the time evolution
of two initially nearby points in the phase space.
Consider the theory with the potential
(\ref{nfldlambda}) with two fields $\phi$ and $\chi$ only
(which we collectively denote as $f$).
Consider two configurations of a scalar field $f$ and
$f'$ that are identical except for a small difference of the
fields at a set of points $x_A$.
Chaos can be defined as the tendency of such
nearby configurations in phase space to diverge exponentially over
time.
This divergence is parametrized by the Lyapunov exponent for
the system, defined as
$\lambda \equiv {1 \over t} log {\Delta(t)\over \Delta_0}$
where $\Delta$ is a distance between two configurations
and $\Delta_0$ is
the initial distance at time $0$.
Numerical results
 shows very
fast onset of chaos around the moment $t_*$ where
the strong turbulence begins.

The highlights of our study for early universe phenomenology are
the following. The mechanism of preheating after inflation is
rather robust and works for  many different systems of
interacting scalars. There is a stage of turbulent classical waves
where the initial conditions for preheating are erased. Initially,
before all the fields have settled into equilibrium with a uniform
temperature, the reheating temperature may be different in
different subgroups of fields. The nature of these groupings is
determined by the coupling strengths.

\section{Conclusion}

   We considered the dynamics of spontaneous symmetry breaking, which
occurs when a scalar field falls down from the top of its effective
potential. We have found  \cite{GBFKLT} that   the main
part of this process typically completes within a single oscillation  of
the distribution of the scalar field. This is a very unexpected conclusion
that may have important cosmological implications.

One of the most efficient mechanisms for the creation of matter after inflation
in theories with convex effective potentials ($V''(\phi) > 0$) is the
mechanism of parametric amplification of vacuum fluctuations in the process
of homogeneous oscillations of the inflaton field, which was called
preheating \cite{KLS}. It has also been noted that in the case where
potentials become concave ($V''(\phi) < 0$), preheating may become more
efficient \cite{Prokopec}. Now we see that this effect is very generic. In
many theories with concave potentials the energy of an unstable vacuum
state is transferred to the energy of inhomogeneous classical waves of
scalar fields  within a single oscillation of the field distribution. We
emphasize here that we are talking about the oscillations of the field
distribution rather than about the oscillations of a homogeneous field
$\phi$ because quite often the homogeneous component $\langle \phi \rangle $ of the field $\phi$
remains zero during the process of spontaneous symmetry breaking.

One of the important consequences of our results is the observation \cite{GBFKLT} that in many models of hybrid inflation  \cite{Hybrid} the first stage of  reheating occurs not due to  homogeneous oscillations of the scalar field  but due to tachyonic preheating \cite{GBFKLT}. 
In particular, this significantly alters the theory of fermionic  preheating \cite{ferm}
for  the hybrid inflation.

The process of preheating and symmetry breaking may take an especially unusual form in the theory of brane inflation \cite{brane} 
based on the hybrid inflation scenario and the mechanism of tachyon condensation on the brane antibrane system \cite{Sen}.

The situation in models of the type used in the new inflation scenario
is somewhat more complicated. In these models the potential is also
concave. However, the expansion of the universe stretches inhomogeneities of
the field rolling down from the top of the effective potential and makes it
homogeneous on an exponentially large scale. Therefore to evaluate a
possible significance of tachyonic instability in this regime one must
compare the amplitude of the homogeneous component of the field with the
amplitude of the quantum fluctuations.  The result appears to be  very
sensitive to the scale  of spontaneous symmetry breaking in such models. A
preliminary  investigation of this issue indicates that in small-field
models where  the scale of spontaneous symmetry breaking  is much smaller
than $M_p$, the leading mechanism of preheating typically is tachyonic. If
correct, this would be a very interesting conclusion  indicating
that in large-field models the leading mechanism of preheating
typically is related to parametric resonance, whereas in small-field models 
 the main mechanism of preheating is typically tachyonic, at least at the first stages of the process.

Finally we should mention that an interesting application of our methods can be found in the recently proposed  ekpyrotic and pyrotechnic scenario \cite{KOST,KKL}. Even though we are very skeptical with respect to the ekpyrotic/pyrotechnic scenario for many reasons explained in \cite{KKL},
 it is still interesting that the methods developed in the theory of tachyonic
 preheating provide us with a very simple theory of the generation of density
 perturbations in these models \cite{KKL}.

I am grateful to Andrei Linde, Gary Felder, Juan J. Garc\'\i a-Bellido Patrick
 Greene   and Igor ~Tkachev for the collaboration on the tachyonic preheating.
 I  thank NSERC, CIAR and the
 NATO Linkage Grant 975389 for support.

\end{document}